\documentclass[11pt,fleqn,twoside]{article}
\usepackage{amsfonts,amssymb,latexsym,epsfig}
\makeatletter
\newcommand{\prava}[1]{\small\it
\begin{flushleft}
Copyright \copyright \ 1999 by  #1
\end{flushleft}}

\newcommand{\name}[1]{\begin{flushleft}
                       \LARGE \bf #1
                       \end{flushleft}\vspace{-3mm}}

\newcommand{\Author}[1]{\begin{flushleft}
                       \it #1 \end{flushleft}}

\newcommand{\Adress}[1]{\begin{flushleft}
                       \it #1 \end{flushleft}}

\newcommand{\Date}[1]{\begin{flushleft}
                      \small  \it #1 \end{flushleft}}

\newcommand{\ehkol}{Author \ name}
\newcommand{\ohkol}{Article \ name}
\renewcommand{\@evenhead}{
\hspace*{-3pt}\raisebox{-15pt}[\headheight][0pt]{\vbox{\hbox to \textwidth 
{\thepage \hfil \ehkol}\vskip4pt \hrule}}}
\renewcommand{\@oddhead}{
\hspace*{-3pt}\raisebox{-15pt}[\headheight][0pt]{\vbox{\hbox to \textwidth 
{\ohkol \hfil \thepage}\vskip4pt\hrule}}}
\renewcommand{\@evenfoot}{}
\renewcommand{\@oddfoot}{}

\newcommand{\be}{\begin{equation}}
\newcommand{\ee}{\end{equation}}
\newcommand{\ba}{\hspace*{-5pt}\begin{array}}
\newcommand{\ea}{\end{array}}

\newcommand{\ds}{\displaystyle}
\makeatother

\begin{document}

\thispagestyle{empty}
\setcounter{page}{314}
\renewcommand{\ehkol}{P. Chanfreau and H. Lyyjynen}
\renewcommand{\ohkol}{Viewing the Ef\/f\/iciency of Chaos Control}

\begin{flushleft}
\footnotesize \sf
Journal of Nonlinear Mathematical Physics \qquad 1999, V.6, N~3,
\pageref{chanfreau-fp}--\pageref{chanfreau-lp}.
\hfill {\sc Article}
\end{flushleft}

\vspace{-5mm}

\renewcommand{\footnoterule}{}
{\renewcommand{\thefootnote}{} \footnote{\prava{P. Chanfreau and H. Lyyjynen}}}

\name{Viewing the Ef\/f\/iciency of Chaos Control}\label{chanfreau-fp}

\Author{Philippe CHANFREAU~${}^\dag$  and Hannu LYYJYNEN~${}^\ddag$}

\Adress{$\dag$~Department of Mathematics, {\AA}bo Akademi, FIN-20500 Turku, Finland \\[1mm]
$\ddag$~Department of Applied Mathematics, Lule\aa \ University of Technology,\\
~~S-97187 Lule\aa, Sweden }

\Date{Received January 19, 1999; Revised April 13, 1999; Accepted
May 14, 1999}

\begin{abstract}
\noindent
This paper aims to cast some new light on controlling chaos
using the OGY- and the Zero-Spectral-Radius methods. In deriving
those methods we use a generalized pro\-cedure dif\/fering from the usual
ones. This procedure allows us to conveniently treat maps to be
controlled
bringing the orbit to both various saddles and to
sources with both real and complex eigenvalues. We demonstrate the
procedure and the subsequent control on a variety of maps. We evaluate
the control by examining the basins of attraction of the relevant
controlled systems graphically and in some cases analytically.
\end{abstract}

\renewcommand{\theequation}{\arabic{section}.\arabic{equation}}

\section{Introduction}\label{intr}

One of the basic properties of a chaotic dynamic system is
topological transitivity. We can take advantage of that property. In
practice it can be used to get from any point in the attractor to the
vicinity of a given point. The popular OGY-method is typically used when
the target point is an unstable f\/ixed point or a point in an unstable
periodic cycle. Once we have arrived
 close enough to the target point we use linear control with respect to
 a parameter in order to stay in that neighborhood. The OGY-method and
 its various versions aim at bringing the orbit to the stable set of the
target point and keeping it there. In that way the system itself will
attract the trajectory to the target. The method is described in
\cite{ogy1} and among many other well formulated descriptions and
derivations we want to mention
 \cite{ogy3,ogy2,dn}. Among others, \cite{h} deals with
 special techniques for stabilising unstable periodic cycles of higher
order. Several other variants of local control through parameter
perturbation and their  comparison to the OGY-method are presented in
\cite{ogl1,ogl3,ogl2}. An investigation of the relations between these
methods and the ``pole placement'' methods can be found for instance
 in \cite{pp}. An attempt to establish a general methodology is made in
\cite{gm3,gm2,gm1}. In order  to get close to a certain point in a
chaotic attractor rapidly from a relatively distant point,various
versions of the ``Targeting''-method can be used. The original idea of
that method is described in \cite{targ1,targ2,targ3}. A large
 overview of early chaos control methods was made in \cite{yle}.
And recently the same authors have assembled most of recent work on the
subject in \cite{bibel}.

In this paper methods of controlling by parameter perturbation in a map
are used to stabilize the system to an unstable f\/ixed point.
A complementary set of examples can be seen in [3].
We will use the OGY-method and the
Zero-Spectral-Radius method. The examples were chosen in order to get a
broad variation in the properties of the unstable f\/ixed points or
periodic cycles. They vary from saddles to sources. Some eigenvalues
of the corresponding Jacobian matrices are real
 and some are complex. We investigate two- and three-dimensional maps.

In some cases control is applied several times on the same map, each
time perturbing a dif\/ferent parameter.

Our main goal is to evaluate the ef\/f\/iciency of the above methods on the
 variety of maps they are tested. The ef\/f\/iciency can be viewed through
the plots (with level curves) of the basins of attraction of the
controlled systems. In the case of the three-dimensional maps the
basins of attraction are viewed in perpendicular cross-sections
through the f\/ixed point. Explicit expressions for the controlled
systems are derived in several cases. In some of the cases the basin
of attraction of the controlled system is so small or so narrow that
the local control method in question is to be employed
with special precaution or to be preceded by a ``targeting'' procedure.

This paper is organized in the following manner:

{\bf Section \ref{nota}.}
The derivation of formulae for linear control by parameter
 perturbation. Both the Zero-Spectral-Radius method (ZSR)
and the OGY-method are derived through matrix manipulations.

{\bf Section \ref{twodim}.}
A study  of the control of  a 2-dimensional discrete map:

The H\'{e}non map is stabilised to a saddle f\/ixed point in several
cases. Control is implemented with respect to several available
parameters and using the two methods mentioned.

{\bf Section \ref{tredim}.}
A study of the control of three-dimensional discrete maps:

In subsection \ref{phattr51}
a H\'{e}non-like map is controlled to a f\/ixed point with one unstable
real eigenvalue and two stable complex ones. Both the OGY- and the
ZSR-methods are used.

In subsection \ref{coupl} a system of three coupled logistic maps
is controlled with the OGY-method. The map is stabilised to a f\/ixed
point with one stable and two unstable directions.

\section{General theory and control formulae}\label{nota}

Let us consider the map
\begin{equation}\label{e1}
\overline{x} \to F(\overline{x},p),
\end{equation}
where $\overline{x}$ is an $n$-dimensional vector and $p$ is a real
parameter.
Without loss of generality we assume the f\/ixed point
$\overline{x}^{\,*}$ to be the origin for the parameter value $p=0$
(if we are interested in other periodic orbits we here use
the corresponding iterate of $f$ instead of $f$ itself).
The map can then be written in the form
\begin{equation}\label{e2}
\overline{x} \to J\overline{x} + p\overline{w} + F_J,
\end{equation}
where $J$ is the Jacobian matrix and $\overline{w} ={{\partial
F}\over{\partial p}}$, both calculated in the f\/ixed
point (here: the origin) and with $p=0$. $F_J$ is of higher order.

To control the system we choose an expression for $p$ being a linear
function of the components of $\overline{x}:$
$p=\overline{\alpha}^T_J\overline{x}$.

Thus the function representing the controlled system
 can also be written in the form
\begin{equation}\label{e3}
\overline{x}\to  J_c\overline{x} +  F_J,
\end{equation}
where $J_c=J+\overline{w}\, \overline{\alpha}^T_J$.

The next step is a change of coordinates.
Suppose the  new coordinates $\overline{u}$ are def\/ined by
$\overline{x}=Q\overline{u}$, where $Q$ is an $n\times n$ non-degenerate matrix.
In the new coordinates the map~(\ref{e1}) can be written in the
following form:
\begin{equation}
\overline{u} \to M\overline{u}+p\overline{b}+ F_M,
\end{equation}
where $M=Q^{-1}JQ, \, \overline{b}=Q^{-1}\overline{w}$ and $ F_M$ is of
higher
order.

The controlled system can be written in the form
\begin{equation}
\overline{u} \to M_c\overline{u} + F_J,
\end{equation}
where $M_c=M+\overline{b} \overline{\alpha}^T_M,\,
\overline{\alpha}^T_M=
\overline{\alpha}^T_JQ$.

 The case where
$M$ becomes a diagonal matrix or a matrix in Jordan normal form
is of special interest.

We shall look at ways of choosing the coef\/f\/icients in order to make
the control work. If the coef\/f\/icients in $\overline{\alpha}_M$
(equivalently
$\overline{\alpha}_J$)
are chosen so that
the absolute value of the eigenvalues are less than one the f\/ixed
point becomes stable for the controlled system. This means that we can
control the system in the vicinity of the f\/ixed point (actually the
controlled system will converge towards  the f\/ixed point).

We shall use two methods:
the Zero-Spectral-Radius method (ZSR) and the OGY-method.
In the ZSR-method all eigenvalues of the Jacobian matrix for the
controlled system are required to be zero. In the OGY-method
only the unstable eigenvalues, that is the ones with absolute
value greater than one, are required to be zero and the
stable ones, those with absolute value less than one are left
unchanged. Let's have a look at the f\/irst case.

\subsection{The Zero-Spectral-Radius method (ZSR)}\label{zsr}

We can derive general expressions for the coef\/f\/icients $\alpha_i$
(the components of $\overline{\alpha}_M$) making all the eigenvalues
of $M_c$ equal to zero.

\medskip

\noindent
{\bf Theorem 2.1.} {\it Let the matrix $M$ be a diagonal matrix
\[
\left(\begin{array}{cccc}
     \lambda_1 & 0 & \cdots & 0 \\
      0 & \lambda_2 & \cdots & 0 \\
     \cdots & \cdots & \cdots & \cdots  \\
      0 & 0 & \cdots & \lambda_n \end{array}\right),
\]
where $\lambda_i\neq\lambda_j$ for $i\neq j$. Furthermore we
suppose that all $b_i\neq 0$. Then the eigenvalues of $M_c$
are all equal to zero if and only if
\begin{equation}\label{coffi}
\alpha_i={{-\lambda_i^n}\over{\prod\limits_{j=1,j\neq i}^n (\lambda_i-
\lambda_j)b_i}},\qquad i=1,\ldots,n.
\end{equation}
($\lambda_i$ can be complex).}

\newpage

\noindent
{\bf Proof.} The characteristic polynomial $P(\lambda )= \det (M_c-\lambda I)
\equiv (-1)^n\lambda^n$, since all its roots are equal to zero.
We will develop $M_c=M+\overline{b} \overline{\alpha}_M^t $.
We get
\[
M_c=\left(\begin{array}{cccc}
     \lambda_1+b_1\alpha_1 & b_1\alpha_2 & \cdots & b_1\alpha_n \\
      b_2\alpha_1 & \lambda_2+b_2\alpha_2 & \cdots & b_2\alpha_n \\
      \cdots & \cdots & \cdots & \cdots  \\
      b_n\alpha_1 & b_n\alpha_2 & \cdots & \lambda_n+
     b_n\alpha_n \end{array}\right),
\]
and $P(\lambda )=\det (M_c-\lambda I)$ equals
\[
\det    \left(\begin{array}{cccc}
     \lambda_1+b_1\alpha_1-\lambda & b_1\alpha_2 & \cdots & b_1\alpha_n \\
      b_2\alpha_1 & \lambda_2+b_2\alpha_2-\lambda & \cdots & b_2\alpha_n \\
      \cdots & \cdots & \cdots & \cdots  \\
      b_n\alpha_1 & b_n\alpha_2 & \cdots & \lambda_n+
     b_n\alpha_n-\lambda \end{array}\right).
\]

Now develop $P(\lambda_1)$ by elementary column operations,
which results in
\[
  \alpha_1\det\left(\begin{array}{cccc}
     b_1 & b_1\alpha_2 & \cdots & b_1\alpha_n \\
      b_2 & \lambda_2-\lambda_1+b_2\alpha_2 & \cdots & b_2\alpha_n \\
      \cdots & \cdots & \cdots & \cdots  \\
      b_n & b_n\alpha_2 & \cdots & \lambda_n-\lambda_1+
     b_n\alpha_n \end{array}\right)
\]
\[
\qquad = \alpha_1\det\left(\begin{array}{cccc}
     b_1 & 0 & \cdots & 0 \\
     b_2 & \lambda_2-\lambda_1 & \cdots & 0 \\
      \cdots & \cdots & \cdots & \cdots  \\
     b_n & 0 & \cdots & \lambda_n-\lambda_1
      \end{array}\right)
 \]
\[
\qquad = b_1\alpha_1(-1)^{n-1}(\lambda_1-\lambda_2)(\lambda_1-\lambda_3)\cdots
     (\lambda_1-\lambda_n)=(-1)^n\lambda_1^n.
\]

Solving for $\alpha_1$ we get (\ref{coffi}) for $i=1$. For
$\alpha_i$, $i=1, \ldots,n$ we get the analogous expressions as
necessary conditions.

If the characteristic polynomial has coef\/f\/icients
$c_i$, $i=0,\ldots,n$, that is $P(\lambda )=c_n\lambda^n+\cdots+c_1\lambda
+c_0$,
then we can immediately see from the determinant expansion that the
leading
coef\/f\/icient must be $(-1)^n$ then giving a system of $n$ equations
$c_{n-1}\lambda_i^{n-1}+\cdots+c_1\lambda_i+c_0=0$, $i=1,\ldots,n$
which has only the trivial solution because the $\lambda_i$:s are
all supposed to be dif\/ferent from each other.
(The system can be written in the form
$C(c_{n-1},\ldots,c_1,c_0)^T=\overline{0}^T$ and it is  well known that
in this case the determinant of $C$ is dif\/ferent from zero).
Thus the conditions are also suf\/f\/icient. The proof is f\/inished.

We shall now look for the concrete formulae in some
special cases used in this paper.

Suppose $J$ is two-dimensional and has two real eigenvalues
$\lambda_1\ne\lambda_2$. Then we get the following
formulae for the coef\/f\/icients:

\begin{equation}\label{2r}
\alpha_1={{-\lambda_1^2}\over{b_1(\lambda_1-\lambda_2)}}, \qquad
\alpha_2={{-\lambda_2^2}\over{b_2(\lambda_2-\lambda_1)}}.
\end{equation}

Suppose $J$ is two-dimensional and has two complex eigenvalues
$\lambda\pm i\mu$, where $\mu\neq 0$.

We remark that
\[
S^{-1}\left(\begin{array}{cc}
     \lambda +\mu i & 0 \\
      0 & \lambda -\mu i \end{array}\right) S=
\left(\begin{array}{cc}
     \lambda  & \mu \\
      -\mu & \lambda \end{array}\right),
\]
where
\[ S=\left(\begin{array}{cc}
     i & 1 \\
      1 & i \end{array}\right).
\]

Thus the expressions for $\alpha_1$ and $\alpha_2$ in this case
 can be obtained from the case with two dif\/ferent real
eigenvalues if we put
$\lambda_1=\lambda +\mu i$, $ \lambda_2=\lambda -\mu i$
and use $\overline{b} =S\tilde b$ and
$\tilde\alpha^T=\overline{\alpha}^TS$,
where $\overline{b}$ gives $b_1$ and $b_2$ for the real eigenvalue case
and  $\tilde b$ gives $b_1$ and $b_2$ for the complex eigenvalue case
and analogously for $\overline{\alpha}$ and $\tilde\alpha$.

Consequently if $M$ is of the form
\[
\left(\begin{array}{cc}
     \lambda & \mu \\
      -\mu & \lambda \end{array}\right),
\]
where $\mu\neq 0$ then we get the following expressions
for the coef\/f\/icients:
\begin{equation}\label{2i}
\alpha_1=(b_2(\mu^2-\lambda^2)-2b_1\mu\lambda)/A, \qquad
\alpha_2=(b_1(\lambda^2-\mu^2)-2b_2\mu\lambda)/A,
\end{equation}
where $A=\mu (b_1^2+b_2^2)$.

Suppose now $J$ is three-dimensional and has one real
eigenvalue $\lambda_u$ and
two complex eigenvalues
$\lambda\pm i\mu$, where $\mu\neq 0$.
 Suppose $M$ is of the form
\[
\left(\begin{array}{ccc}
    \lambda_u & 0 & 0 \\
     0& \lambda & \mu \\
     0&  -\mu & \lambda \end{array}\right),
\]
where $\mu\neq 0$.
Then we get the coef\/f\/icients:
\begin{equation}\label{1r2i}
\alpha_1=-\lambda_u^3/(Db_1),\qquad \alpha_2=-(Bb_2+Cb_3)AD, \qquad
\alpha_3=(Cb_2-Bb_3)/AD,
\end{equation}
where
\[
\ba{l}
A=\mu (b_2^2+b_3^2),\qquad  B=2\lambda^3\mu +\lambda_u\mu^3+2\mu^3\lambda -
3\lambda_u\lambda_2\mu,
\vspace{1mm}\\
C=\lambda^4-\lambda_u\lambda^3+3\lambda_u\mu^2\lambda -\mu^4, \qquad
 D= \lambda^2 +\mu^2-2\lambda_u\lambda+\lambda_u^2.
\ea
\]

\subsection{The OGY formulae}\label{sogy}

The same  techniques and the formulae in the previous
subsection can also be used for the OGY-control. We give the result in some cases.

Suppose for a two-dimensional system the f\/ixed point is a saddle
and $M$ has the form
\[
\left(\begin{array}{cc}
     \lambda_1 & 0 \\
      0 & \lambda_2 \end{array}\right),
\]
where the absolute value of $\lambda_1$ is less than one and
the absolute value of $\lambda_2$ is greater than one.
Then we get the coef\/f\/icients:
\begin{equation}\label{ogf1}
\alpha_1=0,\qquad \alpha_2=-\lambda_2/b_2.
\end{equation}

This is easily seen to coincide with the known form of the
OGY-formula. We have $\overline{\alpha}_J^T=\overline{\alpha}_M^TQ^{-1}$.
And since the columns of $Q$ consist of the eigenvectors of $J$
the rows of $Q^{-1}$ consist of the contravariant basis vectors
$\overline{f}_1$ and $\overline{f}_2$. This leads to
$\overline{\alpha}_J^T=
-\lambda_2\overline{f}_2^T/b_2$. But because
$\overline{b}=Q^{-1}\overline{w}$
we get $b_2=\overline{f}_2\cdot \overline{w} $ resulting in the familiar
\[
\overline{\alpha}_J^T= -{{\lambda_2\overline{f}_2^T}\over{\overline{f}_2^T\overline{w}}}.
\]

Suppose for a three-dimensional system the f\/ixed point
is a saddle with two-dimensional unstable manifold and
$M$ has the form
\[
\left(\begin{array}{ccc}
     \lambda_1 & 0 & 0\\
      0 & \lambda_2 & 0  \\
     0 & 0 & \lambda_3 \end{array}\right),
\]
where the absolute values of $\lambda_1$ and $\lambda_2$
are greater than one and the absolute value of
$\lambda_3$ is less than one.

Then we get the coef\/f\/icients:
\begin{equation}\label{ogf2}
\alpha_1={{-\lambda_1^2}\over{b_1(\lambda_1-\lambda_2)}},\qquad
\alpha_2={{-\lambda_2^2}\over{b_2(\lambda_2-\lambda_1)}},\qquad  \alpha_3=0
\end{equation}
similar to formula (\ref{2r}).

\section{Comparing ways of controlling a  two-dimensional map}\label{twodim}
\setcounter{equation}{0}

We have used our method to control some two-dimensional maps
with only linear and quadratic terms. The f\/ixed points were saddles
and even sources. In this section, however, we will concentrate on
 showing dif\/ferent ways of controlling the familiar H\'{e}non map
with a saddle. The cases with sources are investigated in \cite{preprinten}
and shortly treated in the discussion.

\subsection{The H\'{e}non map}\label{henon}

Let us now consider the map
\begin{equation}\label{h3p1}
        \left( \begin{array}{c} X  \\
                          Y \end{array} \right) \to
\left(  \begin{array}{c}
a-bY-cX^2 \\
   dX+e\end{array} \right).
\end{equation}

For $c=d=1$, $e=0$ this is the famous H\'{e}non map.
The parameters $c$, $d$ and $e$ are introduced in order
to apply control in several ways to the standard
H\'{e}non map. Control is applied by perturbing one
of the parameters while the other ones are held f\/ixed.
For instance applying control with respect to the parameter
$a$ is carried out by permitting the parameter to vary about a nominal
value $a^*$, so that $a=a^*+p$. Applying control with respect to any
of the parameters $b$, $c$, $d$ and $e$ is carried out analogously
by adding the perturbation $p$ to the nominal value of the parameter
in turn to be used. In addition we apply control by adding multiples of
$p$ to both coordinates.

In the following, $c^*=d^*=1$ and $e^*=0$ will be used.
We will assume that the values of $a^*$ and $b^*$ are such that
the map has a chaotic attractor and that within the attractor there is
a saddle f\/ixed point $(x^*,y^*)$, with one stable direction with the
stable eigenvalue~$\lambda_s$ and one unstable direction with
the unstable eigenvalue $\lambda_u$. Notice that, given the above
values of the parameters, $ x^*=y^*. $

One example of such a case is the common version of the H\'{e}non map
with $a^*=1.4$ and  $b^*=-0.3$. It is treated in many papers and
textbooks. In the numerical example in Section~\ref{h3p1p7} we will consider a special case with
values $a^*=1.05$ and $b^*=-0.5$. Iterating the map yields a familiar
multi-folded horseshoe-shaped attractor shown in Fig.~\ref{henfig}{\it a}).
One reason for choosing this particular version of the H\'{e}non map is
that the stable eigenvalue of the f\/ixed point is larger (0.2665) than in
the ``common'' case (0.1559). This fact leads to bigger dif\/ferences in the
results when we apply the two control methods, ZSR and OGY.

We control the system by the ZSR and OGY methods. In order
to use the formulae in Section~\ref{nota} we transform the
system in two steps:

\vspace{-2mm}

\begin{enumerate}

\item[1)] Moving the f\/ixed point to the origin. Separating the linear and non-linear parts.

\vspace{-2mm}

\item[2)] Taking the eigenvectors as new basis vectors.
\vspace{-2mm}
\end{enumerate}

\subsubsection{Moving the f\/ixed point and separating matrix terms}
\label{h3p1p1}

We introduce the new variables $(x,y)=(X-x^*,Y-y^*)$.
The map (\ref{h3p1}) can now be rewritten as
\begin{equation}\label{h3p2}
        \left( \begin{array}{c} x  \\
                          y \end{array} \right) \to
\left(  \begin{array}{c}
a-b(y+y^*)-c(x+x^*)^2-x^* \\
   d(x+x^*)+e-y^*\end{array} \right).
\end{equation}

We apply control with respect to each of the parameters
$a$, $b$, $c$, $d$ and $e$ one by one as described in the
beginning of Section~\ref{henon}.
Expanding and separating the matrix terms
(keeping in mind that $x^*=y^*$, $c^*=1$, $d^*=1$ and $e^*=0$), we
obtain the formula (\ref{e2}) in each case with the following
expression for $J$:
\[
J=\left( \begin{array}{cc} -2x^* & -b^* \\
                  1 & 0 \end{array} \right).
\]

The expressions for $\overline{w}$ will vary:
\begin{equation}\label{w1a}
a) \ \overline{w}=\left(\begin{array}{c} 1\\ 0 \end{array}\right),
\qquad
b) \ \overline{w}=\left(\begin{array}{c} -y^*\\ 0
\end{array}\right),
\end{equation}
\begin{equation}\label{w1b}
c) \ \overline{w}=\left(\begin{array}{c} -(x^*)^2\\ 0
\end{array}\right),
\qquad
d) \ \overline{w}=\left(\begin{array}{c} 0\\ x^* \end{array}\right)
\qquad \mbox{or}\qquad e) \ \overline{w}=\left(\begin{array}{c} 0\\ 1 \end{array}\right),
\end{equation}
depending on which parameter to be perturbed.

In the following we consider especially the f\/irst two cases {\it a}) and {\it b}).
On the other hand we have applied control perturbing both $a$ and $e$
at the same time in the following cases (meaning $(a,e) \to (a,e)^T+p\overline{w}$,
and control with respect to $p$):
\begin{equation}\label{w2}
f) \ \overline{w}=\left(\begin{array}{c} 1\\ -1 \end{array}\right),
\qquad
g) \ \overline{w}=\left(\begin{array}{c} 1\\ 1 \end{array}\right),
 \qquad \mbox{and}\qquad
h) \ \overline{w}=\left(\begin{array}{c} 2\\ 1 \end{array}\right).
\end{equation}

Out of these, we especially consider the case {\it f}).

The expressions for $F_J$ is the same in cases {\it a}), {\it e}), {\it f}), {\it g}) and
{\it h}):
\[
\overline{F}_J=\left(\begin{array}{c} -x^2 \\ 0 \end{array} \right).
 \]

The expression for $F_J$ in case {\it b}) is
\[
\overline{F}_J=\left(\begin{array}{c} -py-x^2 \\ 0 \end{array} \right).
\]

\subsubsection{Changing basis vectors}\label{h3p1p2}

The eigenvalues of $J$ are $\lambda =-x^*\pm\sqrt{(x^*)^2-b^*}$
which results in $\lambda_1\lambda_2=b^*$ and $\lambda_1+\lambda_2
=-2x^*$. As mentioned in the beginning of this section  we are particularly
interested in the f\/ixed point being a saddle. The eigenvalues
can then be noted $\lambda_s$ and $\lambda_u$ and
the eigenvectors are
\[ \overline{e}_s=\left(\begin{array}{c} \lambda_s \\ 1
\end{array}\right),
\qquad\mbox{and}\qquad
\overline{e}_u=\left(\begin{array}{c} \lambda_u \\ 1\end{array}\right).
\]

 The basis is changed so that the eigenvectors of the
Jacobian matrix become the new basis vectors.
The coordinates in the new eigenvector basis are called
$(u,v)$, with $u$ being the coordinate in the stable direction and $v$ the coordinate
in the unstable direction. The matrix $Q$ is the transition matrix
\[
Q = \left( \begin{array}{cc} \lambda_s & \lambda_u \\
                1 & 1 \end{array} \right)
\]
leading to a diagonal matrix $M=Q^{-1}JQ$. Since
$\overline{u} =Q^{-1}\overline{x}$, we get
\begin{equation}\label{h3p4}
u={{1}\over{\lambda_s-\lambda_u}} (x-\lambda_uy), \qquad
v={{1}\over{\lambda_s-\lambda_u}} (-x+\lambda_sy).
\end{equation}

Furthermore, since
\[
\left( \begin{array}{c}
 b_1 \\ b_2\end{array}\right) = Q^{-1}\overline{w},
\]
we get for control with respect to $a$
\begin{equation}\label{h3p5a}
b_1={{1}\over{\lambda_s-\lambda_u}}, \qquad
b_2={{-1}\over{\lambda_s-\lambda_u}},
\end{equation}
we get for control with respect to $b$
\begin{equation}\label{h3p5b}
b_1={{-y^*}\over{\lambda_s-\lambda_u}}, \qquad
b_2={{y^*}\over{\lambda_s-\lambda_u}},
\end{equation}
and for control in the {\it f})-case:
\begin{equation}\label{h3p5f}
b_1={{1+\lambda_u}\over{\lambda_s-\lambda_u}}, \qquad
b_2={{-1-\lambda_s}\over{\lambda_s-\lambda_u}}.
\end{equation}

\subsubsection{Using Zero-Spectral-Radius control} \label{h3p1p3}

Although the f\/ixed point in our H\'{e}non attractor
typically is a saddle point it does not matter in the case
of the ZSR-method if there is a stable eigenvalue or not.
So we will here use the more general notation
$(\lambda_1, \lambda_2)$ instead of $(\lambda_s,\lambda_u)$.
We can now obtain expressions for $\alpha_1$ and $\alpha_2$
in accordance with  (\ref{2r}) leading to explicit expressions
for the parameter perturbation $p=\alpha_1u+\alpha_2v$ for the
H\'{e}non map (using $\lambda_1\lambda_2=b$ and
$\lambda_1+\lambda_2=-2x^*$)

In the case of control with respect to $a$ we get
\begin{equation}\label{h3p6a}
\alpha_1=-\lambda_1^2 , \qquad
\alpha_2=-\lambda_2^2 \qquad \mbox{and}\qquad p=2x^*x+b^*y.
\end{equation}

In the case of control with respect to $b$ we get
\begin{equation}\label{h3p6b}
\alpha_1={{\lambda_1^2}\over{y^*}},
\qquad \alpha_2={{\lambda_2^2}\over{y^*}} \qquad
\mbox{and}\qquad p=-2x-{{b^*}\over{y^*}}y.
\end{equation}

In the case {\it f}) we get
\begin{equation}\label{h3p6f}
\alpha_1={{-\lambda_1^2}\over{1+\lambda_2}}, \qquad
\alpha_2={{-\lambda_2^2}\over{1+\lambda_1}}.
\end{equation}

In the case {\it f}) the analytical expression for $p$ cannot be brought to
a simple form, so it is left out. Thus in the following only the
{\it a})- and {\it b})-cases are treated analytically.

\subsubsection{Deriving the controlled system for the ZSR-method}\label{h3p1p4}

Control parameters are determined from the linear part
of the system, but nevertheless the non-linear part
is still present and together with the  parameter perturbation
a new system, the ``controlled system'' is created.
In general, the controlled system will dif\/fer more from the original
one the further one gets from the f\/ixed point. We insert
the expressions for $p$ into equation (\ref{e2}) and get the following
controlled systems.

\begin{itemize}
\item
In the case of control with respect to the parameter $a$:
\begin{equation}\label{h3p7a}
        \left( \begin{array}{c} x  \\
                          y \end{array} \right) \to
\left(  \begin{array}{c}
-x^2 \\
   x\end{array} \right).
\end{equation}

\item
In the case of control with respect to the parameter $b$:
\begin{equation}\label{h3p7bb}
        \left( \begin{array}{c} x  \\
                          y \end{array} \right) \to
\left(  \begin{array}{c}
\ds -x^2-{{(\lambda_1+\lambda_2)}\over{y^*}}xy+
{{\lambda_1\lambda_2}\over{y^*}}y^2\vspace{2mm}\\
   x\end{array} \right),
\end{equation}
which is the same as
\begin{equation}\label{h3p7b}
        \left( \begin{array}{c} x  \\
                          y \end{array} \right) \to
\left(  \begin{array}{c}
\ds -x^2+2xy+ {{b^*}\over{y^*}}y^2\vspace{2mm}\\
   x\end{array} \right).
\end{equation}
\end{itemize}

In the f\/irst case one can clearly see that the controlled system
is only dependent on $x$ and convergence to the f\/ixed point
(the origin) will occur under the same circumstances  as for the
one-dimensional map $x\to -x^2$. So there is convergence for
(\ref{h3p7a}) to  the f\/ixed point $(0,0)^T$, that is
$(X,Y)^T =(x^*,y^*)^T$, from starting points
with  $X\in\rbrack x^*-1,x^*+1\lbrack$.
For $X=x^*\pm 1$ there is convergence to the point
$(x^*-1,x^*-1)^T$ and divergence otherwise.

\subsubsection{Using OGY-control}\label{h3p1p5}

Since the OGY-method is based on the very property of a f\/ixed point
being a saddle with at least one stable direction we will here
use the notations $(\lambda_s, \lambda_u)$ instead of $(\lambda_1,
\lambda_2)$.

According to (\ref{ogf1}) we get $p=\alpha_1u+\alpha_2v=
0u-{{\lambda_u}\over{b_2}}v$. Thus
\begin{equation}\label{h3p8}
p={{-\lambda_uv}\over{b_2}}.
\end{equation}

Using (\ref{h3p4}) and  (\ref{h3p5a})--(\ref{h3p5f})
we can get the particular expressions for each version of
parameter perturbation.

\subsubsection{Deriving the controlled system for the OGY-control}\label{h3p1p6}

Inserting the expressions for $p$ into equation (\ref{e2}) we get
the following  expressions for the controlled system applying OGY-control.

\begin{itemize} \item
In the case of control with respect to $a$:
\begin{equation}\label{h3p9a}
        \left( \begin{array}{c} x  \\
                          y \end{array} \right) \to
\left(  \begin{array}{c}
\lambda_sx-x^2 \\
   x\end{array} \right).
\end{equation}

\item
In the case of control with respect to $b$:
\begin{equation}\label{h3p9b}
        \left( \begin{array}{c} x  \\
                          y \end{array} \right) \to
\left(  \begin{array}{c}
\ds \lambda_sx-x^2-{{\lambda_u}\over{y^*}}xy+{{\lambda_s\lambda_u}\over{
y^*}}y^2\vspace{2mm}\\
   x\end{array} \right).
\end{equation}
\end{itemize}

Here again in the f\/irst case the controlled system is only dependent on $x$ and
convergence to the f\/ixed point (the origin) will occur under the same circumstances
as for the one-dimensional map $x\to \lambda_sx-x^2$.
 So there is convergence for (\ref{h3p9a}) to the f\/ixed
point $(0,0)^T$, that is  $(X,Y)^T = (x^*,y^*)^T$,
 from starting points with  $X\in \rbrack x^*+\lambda_s-1,x^*+1\lbrack$.
  For $X=x^*+\lambda_s-1$ and $X=x^*+1$   there is convergence to the point
  $(x^*+\lambda_s-1,x^*+\lambda_s-1)^T$  and divergence otherwise.

\begin{figure}[t]

\vspace*{2mm}

{\centerline{
\epsfxsize=145mm
\epsfbox{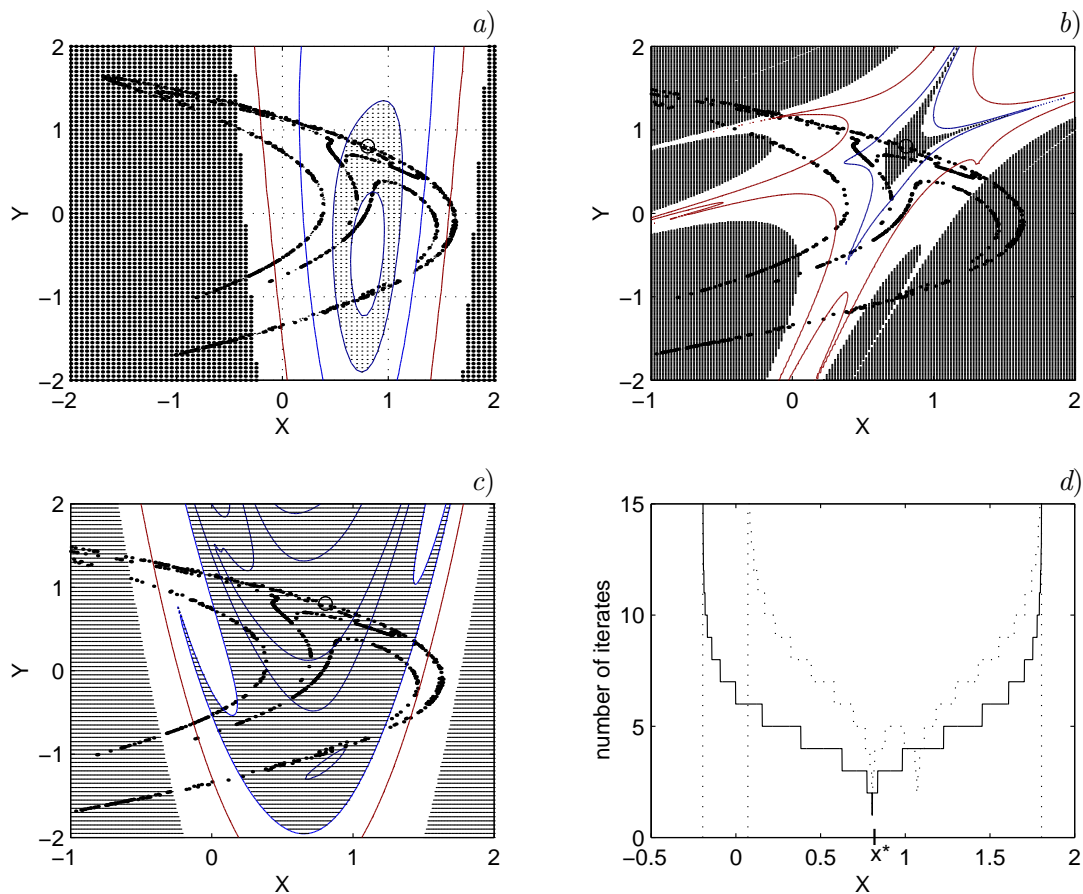}}}
\begin{picture}(0,0)(0,0)
\put(65,120){{\it a})}
\put(143,120){{\it b})}
\put(65,59){{\it c})}
\put(143,59){{\it d})}
\end{picture}

\vspace{-6mm}

\caption{{\it a}) OGY-control in the case  {\it f}. {\it b}) OGY-control with respect to
the parameter $b$. {\it c}) ZSR-control in the case $f$.
{\it d}) OGY- and ZSR-control with respect to the parameter {\it a}.} \label{henfig}

\vspace{-3mm}

\end{figure}

\subsubsection{A numerical example}\label{h3p1p7}

The H\'{e}non map with parameter values $a = 1.05$ and $b = - 0.5$,
(and $c = d =1$, $e = 0$) will be used. The corresponding chaotic attractor is
shown in Fig.~\ref{henfig}{\it a}). The unstable f\/ixed point is $(0.8048, 0.8048)^T$
and
it is marked with a circle (o) in the plot. The eigenvalues for the
Jacobian matrix are $\lambda_s =$ 0.2665 and $\lambda_u =$ -1.8760. The
corresponding eigenvectors are $\overline{e}_s = (0.2665, 1)^T$
and $\overline{e}_u = (-1.8760, 1)^T$.
The two methods, OGY and ZSR, yield almost congruent basins of
attraction using the same parameter      (parameters $b$, $c$, $d$ or $e$). On the
other hand depending on the parameter to be perturbed the basins of attraction
are very dif\/ferent in shape. In the cases {\it g}) and {\it h}) the basins are
similar in shape and geometry regardless of the method (mostly almost vertical
contour lines). The {\it f})-case application of the
two methods yields a vastly dif\/fering inner geometry, as can be seen in the
Fig.~\ref{henfig}{\it a}) and~\ref{henfig}{\it c}).
In all cases the ZSR-method displays larger  regions of the
fastest convergence than the OGY-method perturbing the same parameter.

In Fig. {\it a}), {\it b}) and {\it c}) the map is iterated 5 times. The central
dotted regions contain the points from which the map converges and reaches
within a distance of less than  0.001 from the f\/ixed point (o). The
peripheral dotted regions represent points from which the map certainly
diverges.

Fig.~{\it a}) displays the result of  OGY-control  in the {\it f})-case. The level
curves are: 0.001, 0.01 and 0.1.
Fig.~{\it b}) shows the result of OGY-control with respect to the parameter~$b$.
The level curves are: 0.003 and 0.06.
Fig.~{\it c}) displays the result of ZSR-control in the {\it f})-case. The level
curves are: 0.00001, 0.0001, 0.001 and 0.01.
Finally in Fig.~{\it d})  the results of control using both OGY- and
ZSR-methods with respect to the parameter $a$  are displayed. It was
stated in subsections~\ref{h3p1p4} and
\ref{h3p1p6} that the controlled systems only depend on $x$. The graphs
show the number of iterates needed to get within a distance of
0.001 from the f\/ixed point from various starting points. The solid graph
represents the ZSR-method. The dotted lines $X = x^*\pm 1$ mark the
boundaries of the basin of attraction.
The dotted graph represents the OGY-method. The
dotted lines $X = x^*+\lambda_s -1$ and
$X = x^* + 1$ mark the boundaries of the basin of attraction.

In this section we have not so far paid particular interest in limiting
the size of the parameter perturbation  $p$. We have looked for the
entire basin of attraction regardless of how much an unlimited value for
$p$  might distort the system. This calls for a discussion. Where is the
region of the smallest initial $p$:s?  Let us have a look at the
OGY-method f\/irst. The parameter perturbation  $p$ is generally expressed
in formula~(\ref{h3p8}). Clearly the smallest values for  $p$  are required
close to the straight line determined by  $v = 0$ (formula (\ref{h3p4})),
that is along the direction of the stable eigenvector $(\lambda_s, 1)^T$
or $(1, 3.75)^T$. This applies in all of our cases
and is no big surprise giving the basic idea of the OGY-method.
Investigating the expressions for $p$ (formulae (\ref{h3p6a})--(\ref{h3p6b})
and the other cases {\it c})--{\it h}) not explicitly given in this text)
in the ZSR-method yield correspondingly straight lines as the
centre lines  of stripes constituting the regions of the smallest
initial $p$:s. These lines have directions $(1, -2x^*/b^*)^T$  or
$(1, 3.22)^T$ in the cases {\it a}), {\it b}) and {\it c}). The cases {\it d}) and~{\it e}) have a
centre line with direction $(1,3.84)^T$. In many cases the direction
of the line f\/its well in the image of the basins of attraction. For instance, in Fig.~\ref{henfig}{\it b})
this stripe of small $p$:s follows narrow parts of fast convergence in the basin of attraction.
Nevertheless, the further one gets along the line from the f\/ixed
point the more powerful the non-linear part of the controlled system
becomes yielding uncertain results.

\section{Controlling three-dimensional maps} \label{tredim}
\setcounter{equation}{0}

\subsection{Controlling a 3-dimensional H\'{e}non-like map}\label{phattr51}

Let us consider a three-dimensional map somewhat similar to the
H\'{e}non map:
\begin{equation}\label{e511}
\left( \begin{array}{c} x \\ y \\ z\end{array} \right)
\to
\left( \begin{array}{c} ax+by+cz-x^2+p \\ x \\ y\end{array} \right),
\end{equation}
where normally p = 0.

We will investigate the ef\/fect of controlling the map with the
OGY-method and the Zero-Spectral-Radius method. We will
consider a case where there is a chaotic attractor including an
unstable f\/ixed point (the origin) with one unstable direction,
with the eigenvalue $\lambda_u$, and a stable manifold with two complex
eigenvalues $\lambda_2$ and $\lambda_3$.
We will apply control using the parameter $p$.

\subsubsection{The control formulae}

In order to write the OGY-control formula
we refer to Section~\ref{nota} formula~(\ref{ogf1}).
Analogously with that formula we have $\alpha_1 =-\lambda_u /b_1$,
$\alpha_2 = 0$ and $\alpha_3 = 0$
yielding  $p=-\lambda_u u_1/b_1$, where $u_1$ is the f\/irst
coordinate of the new basis vector $\overline{u}$.

As to the ZSR-control formula we refer to Section \ref{nota} formula~(\ref{1r2i}).
The derivation of the coef\/f\/icients is quite tedious but is conveniently
carried out with the aid of a computer.

\subsubsection{The controlled systems}

The controlled system of the map applying the control methods
can be derived as follows. Consider the Jacobian matrix of the
controlled system:
\[
\left( \begin{array}{ccc} A & B & C \\
                  1 & 0 & 0 \\
                  0 & 1 & 0 \end{array} \right),
\]
where $A = a + \alpha_1$, $B = b + \alpha_2$, $C=c+ \alpha_3$ and
$p= \alpha_1x + \alpha_2 y + \alpha_3 z$.
The characteristic equation is
\begin{equation}\label{e512}
    \lambda^3-A\lambda^2-B\lambda -C=0.
\end{equation}

 In the case of  OGY-control  the characteristic equation
  can also be  written in the following form since the one
  unstable eigenvalue is required to be zero:
  $\lambda (\lambda -\lambda_2)(\lambda -\lambda_3)=0$.
 Developing and
  comparing the two equations yield:
  $A=\lambda_2+\lambda_3$, $B=-\lambda_2\lambda_3$, and  $C = 0$. So the
map has a controlled system of the  following form when $\lambda_u$ is
  required to be zero in accordance with the OGY-method:

\begin{equation}\label{e513}
\left( \begin{array}{c} x \\ y \\ z\end{array} \right)
\to
\left( \begin{array}{c} (\lambda_2+\lambda_3)x-
\lambda_2\lambda_3y-x^2 \\ x \\ y\end{array} \right).
\end{equation}

  This implies that the controlled system does not depend on $z$.

  In the case of ZSR-control the characteristic equation can
  also be written in the following form since all eigenvalues
  are required to be zero:  $\lambda^3 = 0$.

Comparing with the equation  \ref{e512} yields:
  $A = B = C = 0$. So the linear part of the f\/irst equation vanishes and
  the map has a controlled system of the  following form:
\begin{equation}\label{e514}
\left( \begin{array}{c} x \\ y \\ z\end{array} \right)
\to
\left( \begin{array}{c} -x^2 \\ x \\ y\end{array} \right).
\end{equation}

As for the H\'{e}non map the map \ref{e514} will depend only on  $x$
and  convergence will occur under the same circumstances as
the one-dimensional map  $x \to -x^2$. That is the map will
converge to the f\/ixed point $(0, 0, 0)^T$ from any point with
$x\in\rbrack -1, 1\lbrack$.
For points with $x = \pm 1$  there is convergence to
the point $(-1,-1,-1)^T$ and divergence otherwise.

\subsubsection{A numerical example}

We will consider the particular case:
\begin{equation}\label{ex51}
\left( \begin{array}{c} x \\ y \\ z\end{array} \right)
\to
\left( \begin{array}{c} -1.65x-0.3y-0.2z-x^2+p \\ x \\ y\end{array}
\right).
\end{equation}

\begin{figure}[t]
\vspace*{2mm}

{\centerline{\epsfxsize=145mm
\epsfbox{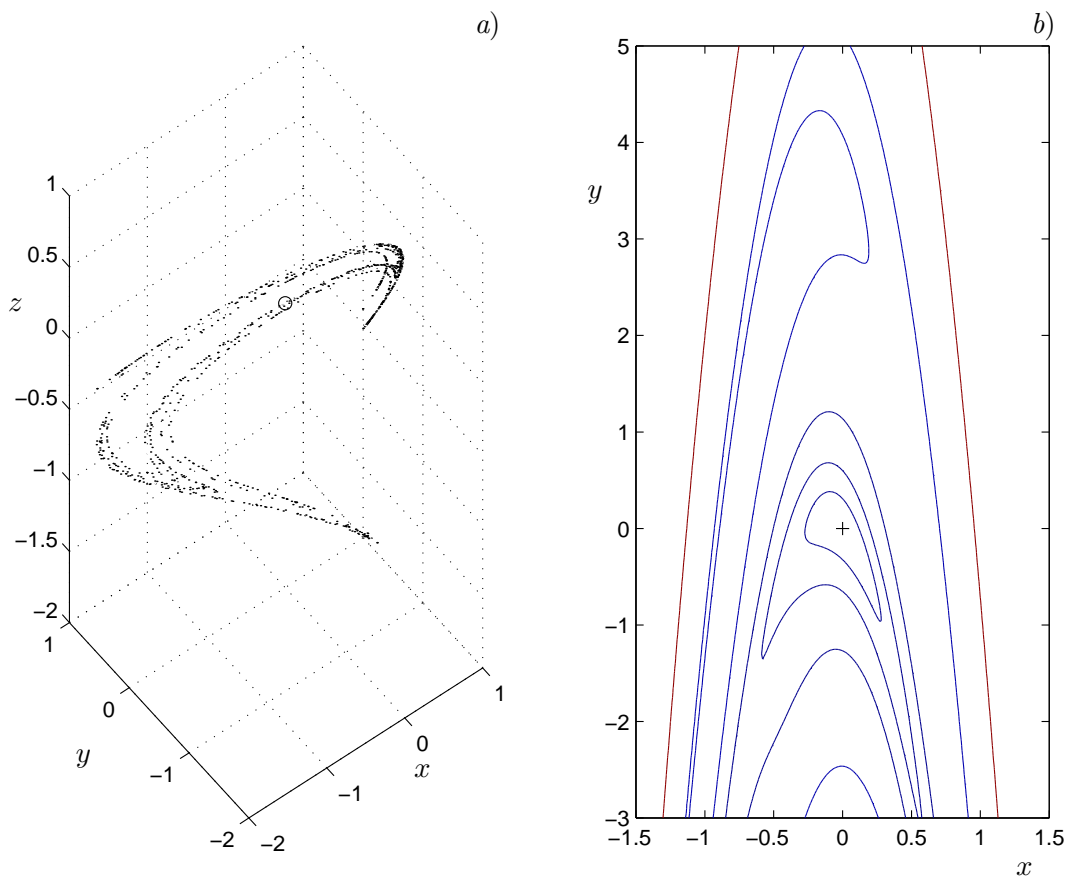}}}
\begin{picture}(0,0)(0,0)
\put(68,117){{\it a})}
\put(142,117){{\it b})}
\put(6,80){$z$}
\put(15,20){$y$}
\put(60,18){$x$}
\put(83,95){$y$}
\put(140,5){$x$}
\end{picture}

\vspace{-5mm}

 \caption{{\it a}) The attractor. {\it b}) A cross-section of the basin of
 attraction through the f\/ixed point at $z=0$ using OGY-control.}\label{phiher}

\vspace{-3mm}

\end{figure}

The corresponding attractor can be seen in Fig.~\ref{phiher}{\it a}).

As can easily be seen the map has a f\/ixed point in the origin
$(0,0,0)$ situated in the chaotic attractor.  The eigenvalues and
eigenvectors of the Jacobian matrix in this point are

\[
\ba{l}
  \lambda_u = -1.5395,
\vspace{1mm}\\
\overline{e}_u= (0.7906,-0.5135,0.3336)^T,
\vspace{1mm}\\
\lambda_{2,3}=-0.0552\pm 0.3562i,
\vspace{1mm}\\
\overline{e}_{2,3}=(-0.1211\mp 0.0076i, -0.0306\pm 0.3352i
-0.9059\mp 0.2265i)^T.
\end{array}
\]

That is, there are one real unstable eigenvalue $\lambda_u$ and two
complex stable eigenvalues $\lambda_2$ and $\lambda_3$.

We shall now have a look at the basin of attraction and the
rate of convergence of  the control methods depending on the
starting point. Three cross-sections were made through the f\/ixed point:
those were the planes $x=0$, $y =0$ and $z=0$. The controlled  system
was iterated f\/ive times starting from each point on each plane.
Contour lines show how close the system point has reached the
f\/ixed point after f\/ive iterations. We choose to display the result only
from the cross-section $z=0$.

{\bf Controlling with the OGY-method.}
Applying the method in order to control the map to the f\/ixed
point  $(0, 0, 0)^T$ works well within certain limits of distance
from that point. First, the planes $x=0$ and $y =0$ only display
parallel contour lines which is explained by the controlled
system not depending on $z$. The contour lines of the plane $z=0$
are shown in Fig.~\ref{phiher}{\it b}).

The f\/ixed point is marked with a cross $(+)$. The vaguely
boomerang shaped contour lines represent the levels (counted
from inside out): 0.005, 0.010, 0.020, 0.050, and 1.000. The last level
represents more or less the boundary of the basin of
attraction for the OGY-controlled system. The 0.020 level makes an extra
turn at the left side.

{\bf Controlling with the Zero-Spectral-Radius method.}
The $y=0$ and $z=0$ planes display parallel lines and $x=0$ does not
display any contour lines at all. This is explained by looking
at the extremely simple expression for the controlled system.

From all starting points in the $x=0$ plane the system goes
exactly to the origin in the second iteration at the latest.~\label{phattr}

\subsection{Coupled logistic map with OGY-control}\label{coupl}

In this subsection we examine the ef\/f\/iciency
of the OGY-control for another three-di\-men\-sio\-nal map. The map is
stabilized towards a saddle f\/ixed point
with two unstable eigendirections. Another version of this map with a
saddle point with
one unstable eigendirection is examined in~\cite{preprinten}.
We consider the system consisting of three coupled logistic
maps
\[
\left( \begin{array}{c}  x \\ y \\ z \end{array} \right) \to
\left( \begin{array}{ccc}  1-2 p & p &  p \\
                        p  & 1-2 p &  p \\
                        p &  p &   1-2 p
\end{array} \right)
\left( \begin{array}{ccc} r_1 & x & (1-x) \\
                        r_2 & y & (1-y) \\
                        r_3 & z & (1-z)
\end{array} \right),
\]
where $r_1$, $r_2$, $r_3$ and $p$ are parameters. The coupling parameter
$p$ is chosen to be the control parameter. As the state space of the
system we consider the unit cube $\lbrack 0,1 \rbrack \times \lbrack 0,1 \rbrack
        \times  \lbrack 0,1 \rbrack$.

        With parameter values $ r_1 = 3.9$, $r_2 = 3.95$,
$r_3 = 1.0 $ and $ p = 0.0736 $, the point
\[
\overline{x}^{\,*T} = (0.7291, 0.7323, 0.2889)
\]
is a f\/ixed point and there are  one stable and two unstable eigendirections.

        The stable eigenvalue and the unstable ones with the
corresponding eigenvectors are
\[
\begin{array}{lcl}
\lambda_s =  0.35600 & \qquad &
        \overline{e}_s =  (0.021204, 0.020760, 1.3728)^T
\vspace{1mm}\\
\lambda_{u1} = -1.4094   &  &
        \overline{e}_{u1} =  (0.63971, -0.54033, 0.0063062)^T
\vspace{1mm}\\
\lambda_{u2} =  -1.6750 &  &
        \overline{e}_{u2} = (0.57127, 0.65842,  0.080606)^T.
   \end{array}
\]

The vector $\overline{w}$ becomes
\[
 \overline{w} = \frac{\partial f}{\partial p}(\overline{x}^{\,*},p^*)  =
 (-0.56084,   -0.57308,   1.1339 ).
\]

Here we  calculate the control coef\/f\/icients according to
(\ref{ogf2}) using again in Section~\ref{sogy}
$\overline{x}-\overline{x}^*$ and $p-p^*=\delta p$
instead of $\overline{x}$ and $p$.

        The explicit expression for the control is obtained as
\begin{equation}
\delta p = 104.33 x -107.46 y + 0.013503 z + 2.6172.
\end{equation}

        The controlled map has the Jacobian matrix
\[
{J_c} =  \left(
{\begin{array}{ccc}
-60.036 & 60.130 & 0.023502 \\
-59.921 & 60.016 & 0.023337 \\
118.17  & -121.98 & 0.37538
\end{array}}
\right)
\]
 with $(0.11687,0.1163,1)$ as eigendirection corresponding to the
eigenvalue zero and \linebreak $(1,0.97904,64.737)$ as eigendirection for
the non-zero eigenvalue.

\begin{figure}[t]
\vspace*{2mm}

{\centerline{
\epsfxsize=135mm
\epsfbox{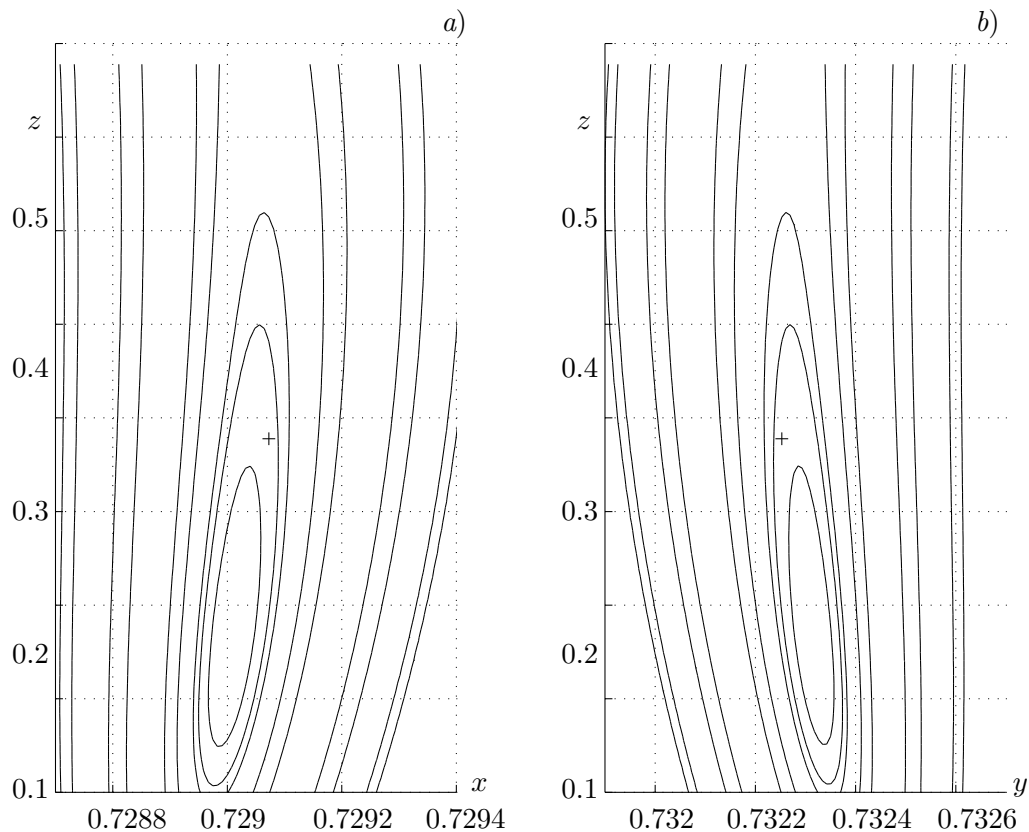}}}
\begin{picture}(0,0)(0,0)
\put(10,85){0.5}  \put(20,5){0.7288}
\put(10,65){0.4}  \put(92,5){0.732}
\put(10,46){0.3}  \put(105,5){0.7322}
\put(10,27){0.2}
\put(10,9){0.1}

\put(83,85){0.5}  \put(119,5){0.7324}
\put(83,65){0.4}  \put(133,5){0.7326}
\put(83,46){0.3}  \put(50,5){0.7292}
\put(83,27){0.2}  \put(65,5){0.7294}
\put(83,9){0.1}   \put(35,5){0.729}
                \put(67,111){{\it a})}
                \put(138,111){{\it b})}
\put(12,98){$z$}   \put(85,98){$z$}
\put(143,10){$y$} \put(71,10){$x$}
\end{picture}

\vspace{-4mm}

\caption{One stable, two unstable eigendirections. Convergence levels
 for {\it a}) $y=0.7323$ and {\it b}) $x=0.7291$.}\label{figh3}
\end{figure}

  The cross-sections through the f\/ixed points of the small region of
convergence are depicted in Fig.~\ref{figh3}.
The convergence levels are (calculated using 10 iterates)
$ (0.048, 0.125, 0.5$, $1, 5, 10, 50, 100) \cdot 10^{-9}.$

\section{Discussion}\label{discu}

The message of this paper is two-fold:
\vspace{-2mm}

\begin{enumerate}
\item[1.] The display of basins of attraction with level-curves for the controlled systems for a
 variety  of maps.
\vspace{-2mm}

\item[2.] A new procedure for deriving control formulae for the  OGY-method and related methods.
\vspace{-2mm}
\end{enumerate}

We have chosen a variety of maps covering some unusual cases
with real or complex eigenvalues. For this purpose
we have brought forward a procedure, which in our opinion simplif\/ies the
derivations of control formulae through matrix diagonalising. Moreover
the procedure yields convenient expressions (Section~\ref{nota}).
The OGY-method and the ZSR-method thereby appear as mere special cases
of a general way of thinking, regardless of the number of dimensions,
the proportion of unstable eigendirections or the quality of the
eigenvalues. There are possibilities to bring forward other similar
methods than the mentioned ones. A comparison with the original
form of the OGY formula is given in subsection~\ref{sogy}. Our procedure is
implemented in all sections but it is displayed more in detail in
Section~\ref{twodim}. This section  \ref{twodim} contains a
broad study of many possible ways to control
the well-known H\'{e}non map. The goal of the control is to bring the
orbit to a saddle f\/ixed point, which works successfully in all cases.
But the geometry of the basin of attraction varies considerably depending
on how the control is applied.

Our ``three-dimensional H\'{e}non" map has two complex
eigenvalues in the origin and serves as an example of applying our
procedure in such a case.

In the case of the three-dimensional coupled logistic map all the
eigenvalues are real, one of them stable and two unstable.
The control coef\/f\/icients become very large and the basin of attraction
to the f\/ixed point very small.  The eigenvector corresponding
to the eigenvalue zero (in the controlled system)  dictates
the direction of fastest convergence.

\strut\hfill

Further examples with convergence level graphs can be found in~\cite{preprinten}:
\begin{itemize}
\item Another version of the coupled logistic map with one unstable and
two stable eigendirections and controlled with the OGY-method.
\item Two-dimensional
maps are stabilised to sources with either real or complex
eigenvalues. Obviously, the OGY-method is not applicable.
The control with the ZSR-method worked well
although the basins of attraction were rather narrow due to the fact
that the two eigenvalues were near each other in size.
This was the case also in Section~\ref{coupl}.
\item
There is also an example on  OGY-controlling the Poincar\'{e} map of a Duf\/f\/ing system
either to a saddle f\/ixed point or to a two-periodic saddle point.
An analysis of the rate of convergence is included.
\end{itemize}

As to the local speed of convergence: the ZSR is always the more
ef\/f\/icient method due to the fact that the eigenvalues of all directions
are placed at zero. The speed of the OGY method depends on the value of the stable
eigenvalues. If their absolute value is close to one, the convergence is
slow. If it is close to zero the behaviour is similar to the one of the
ZSR-method. On the other hand, further away from the f\/ixed point the
importance of the non-linearity of the map grows. There the properties of
the linearisation almost vanish. Therefore, in general, both methods
display almost congruent outer boundaries of their basins of attraction.
The non-linear inf\/luence also create fairly distant areas with fast
convergence requiring only limited control.  Much closer areas may have
slow convergence or even divergence.

We have concentrated on applying linear control to one parameter at a
time. In general we would archive better results by perturbing several
parameters or by applying non-linear control. Another procedure would
be to combine the linear methods with targeting.

\subsection*{Acknowledgements}
The authors are in debt for valuable consultations
to the staf\/f of the  Departments of Mathematics at the
 Lule\aa\  University of Technology and \AA bo Aka\-de\-mi.

\label{chanfreau-lp}

\end{document}